\documentclass[%
 reprint,
 amsmath,amssymb,
 aps,
]{revtex4-2}

\usepackage{graphicx}
\usepackage{dcolumn}
\usepackage{bm}
\usepackage{hyperref}

\usepackage{xcolor}
\usepackage[T1]{fontenc} 
\usepackage[utf8]{inputenc}
\begin{document}

\preprint{APS/123-QED}

\title{Presentation of the electromagnetic field in introductory
  physics textbooks and consequences for its teaching}

\author{Álvaro Suárez}
\affiliation{Departamento de Física, Consejo de Formación en
  Educación, Montevideo, Uruguay}

\author{Arturo C. Marti}
\affiliation{Instituto de F\'{i}sica, Facultad de Ciencias,
  Universidad de la Rep\'{u}blica, Igu\'{a} 4225, Montevideo, 11200,
  Uruguay}

\author{Kristina Zuza}
\affiliation{Department of Applied Physics and Donostia Physics
  Education Research Group, University of the Basque Country
  (UPV/EHU), San Sebastian 20018, Spain}

\author{Jenaro Guisasola}
\affiliation{Donostia Physics Education Research Group, University of
  the Basque Country (UPV/EHU), San Sebastian 20018, Spain \\ School
  of Dual Engineering, Institute of Machine Tools (IMH), Elgoibar,
  Spain}

\date{\today}

\begin{abstract}
Textbooks play a fundamental role in teaching and learning in school
science classrooms. In this paper we investigate the presentation of
the nature of the electromagnetic field in a dozen of the world's most
popular introductory university physics textbooks. We analyse, from an
epistemologically based teaching approach, the didactic treatment of
the electromagnetic field in relation to its sources, Maxwell's laws
and electromagnetic waves. With this objective, we elaborate a
rational reconstruction of the developments that led to the
formulation of the nature of the electromagnetic field, Maxwell's laws
and their meaning, as well as electromagnetic waves. Next, we
formulate criteria based on the key aspects derived from the
reconstruction that are useful in the evaluation of electromagnetism
textbooks at the introductory level and apply them to the sample of
selected books. In light of the results, we reflect on their
consequences for teaching. Our analysis indicates the existence of
certain inconsistencies in the approach to the electromagnetic field
and its relationship with its sources, Maxwell's laws and
electromagnetic waves in many of the books analysed.
\end{abstract}

\maketitle 

\section{Introduction}
The teaching/learning of electromagnetism has been the object of
multiple studies in Physics Education Research (PER)
\cite{galili1997changing,chabay2006restructuring,Zuza2014,jelicic2017analyzing,li2019investigating,campos2021,Campos2023}. Nevertheless,
it still raises significant challenges in many aspects, such as the
attribution of causal relationships between electromagnetic fields and
the difficulty of identifying the nature of their sources
\cite{french2000maxwell,jefimenko2004presenting,hill2010rephrasing,hill2011reanalyzing,Tran_2018,suarez_marti_zuza_guisasola_2022,suarez2022unified}. In
fact, research into high school and university students’ difficulties
in understanding and implementing Maxwell’s laws has systematically
proven that a significant percentage of students still rely on
misconceptions
\cite{rainson1994students,guisasola2008gauss,Manogue,guisasola2013university,wallace2010upper,campos2019}. This
poor learning progress has been attributed to different factors,
including the nature of conventional science teaching and the
difficulties that hinder a better learning of scientific
theories. Empirical research has also established that physics
textbooks play a crucial role when it comes to presenting scientific
theories and models in an accurate and coherent way. There is no doubt
about the need to do further research into the representation, the
logical correction, the structure and the systematics of the contents
in textbooks, regarding specific curriculum topics. In this paper, we
analyse the ways the nature of the electromagnetic field is presented
in a dozen textbooks, which rank amongst those most frequently used to
teach introductory physics at universities.

The analysis of textbooks is justified by the importance of their
influence on both teaching and learning. Research has systematically
established that, to a large extent, science textbooks dictate the
content and emphasis of science study plans, as well as the nature and
scope of instructional activities and the discourse in most classrooms
\cite{chiappetta1991quantitative,shiland1997quantum,roseman2010method}.
Undoubtedly, they are a mayor curricular and didactic resource for
both junior and experienced science teachers. Students also use them
in addition to other online resources to do their homework and prepare
for exams \cite{Ruggieri}. Thus, the structure, the logic and
coherence of the contents is crucial for its educational efficiency
\cite{Korhasan,Zeng}.

International organizations such as UNESCO promote the educational
analysis of textbooks and put forward factors to consider in such
analysis \cite{pingel2010unesco}. The American Association for the
Advancement of Science (AAAS) Project 2061 defends the development of
analysis protocols to evaluate the educational efficiency of science
textbooks \cite{koppal2004meeting}, due to the fact that they
frequently represent over 75 percent of the tasks assigned by the
teacher \cite{chiappetta2006examination}. In this research, we focus
on introductory physics textbooks published in the USA, since they are
the most widely recommended in introductory physics courses in Science
and Engineering degrees worldwide. It can reasonably be supposed that
the trends in American science textbooks are likely to significantly
impact the teaching and learning of sciences in multiple world markets
\cite{butler2009textbook}.  Furthermore, it is reasonable to suggest
that textbook evaluation, based on criteria derived from the key
aspects of rational reconstruction of classical electromagnetism, can
provide teachers with an idea of how models or theories
develop. Ignoring such key aspects and their development in textbooks
can deprive students of the opportunity of becoming familiar with the
scientific practice and progress.

In accordance with the above considerations, our research question is
the following: How do introductory physics textbooks explain the
nature of the electromagnetic field? More specifically, we aim to
answer these research questions:
\begin{itemize}
    \item How do they explain Maxwell's laws and what attention is
      paid to the sources of the fields?
\end{itemize}
\begin{itemize}
    \item How do they treat the generation and propagation of
      electromagnetic waves?
\end{itemize}
\begin{itemize}
    \item How is the electromagnetic field presented in relation to
      different reference systems?
\end{itemize}

In order to answer these questions, we analyse, from an
epistemologically grounded approach to teaching, how physics textbooks
deal with the electromagnetic field in relation to its sources,
Maxwell’s laws and electromagnetic waves. To achieve this objective,
first we carry out a rational reconstruction of the developments that
led to the formulation of the nature of the electromagnetic field,
Maxwell’s laws and their meaning, as well as electromagnetic waves in
classical electromagnetism. Then, we formulate the criteria based on
key aspects derived from the reconstruction, which may yield results
in the evaluation of introductory-level electromagnetism textbooks. We
examine the textbooks and, in the light of results obtained, consider
their implications for teaching.

\section{\label{Theorical}THEORETICAL FRAMEWORK} 

Recent studies that examine science teaching underpin the importance
of both the history and the philosophy of sciences for science
teaching \cite{duschl1994research,matthews2017history}. The
epistemologically grounded approach to teaching involves considering
contributions related to the nature of science and its influence on
teaching
\cite{guisasola2005nature,ruthven2009design,duit2012model,zuza2020towards}. This
approach indicates that knowledge results from a complex process where
problems are solved and initial hypotheses tested, a process that
enables the identification of the way ideas evolve to their current
status \cite{nersessian1995should}. Recognising the evolution of the
concepts underlying the different models can contribute to identifying
epistemological and ontological barriers that scientists have to
overcome in order to develop their theories. Such epistemic
characteristics can usefully inform teaching approaches and help avoid
inaccurate or excessively simplistic perspectives. Furthermore, we
must keep in mind that educational standards set in the last few
decades \cite{national1996national,rocard2007science} demand that the
presentation of concepts and theories is meaningful and includes the
origin and development of key ideas. As a consequence, textbooks are
required to not only focus on the “products” of science,
i.e. concepts, theories, principles and laws of nature, but also put
them into context within the epistemological characteristics that
enabled their development \cite{bensaude2006textbooks}.

Concerning the process of constructing epistemologically grounded
teaching, one major requirement is to detect key features of the
epistemological development of theories the understanding of which can
facilitate students’ learning
\cite{guisasola2014teaching,hodson2014nature}. Scientific research
tends to seek explanations to empirical or theoretical questions that
are not understood, generating research problems that seek to define
key aspects that allow the development and consolidation of theories
\cite{lakatos1970history}. Science curricula cannot ignore the
importance of these key aspects that allowed the development and
consolidation of theories.

The history of electromagnetic theory from the 19th to the late 20th
century shows the models of M. Faraday, J. C. Maxwell, H. Lorentz and
O. Jefimenko evolving in rapid succession. They had to overcome
competing models based on ontological beliefs and rival research
programs, such as the theory of action at a distance or the ether
theory. This period, in which Maxwell's laws present a framework for
analysing electromagnetic interactions and identifying their sources,
has spurred much debate and controversy in the History and Philosophy
of Physics literature
\cite{harman1982energy,whittaker1989history,nersessian2012faraday,berkson2014fields}.
According to this perspective, it is important to analyse introductory
university-level electromagnetism textbooks to determine the extent to
which they address the key aspects of the classical theory of the
electromagnetic field, its fundamental laws, and the sources of the
electromagnetic field.

\section{\label{Epistemology}CONTRIBUTIONS OF EPISTEMOLOGY OF PHYSICS ON THE NATURE OF THE ELECTROMAGNETIC FIELD} 

During the 19th century, the confrontation between two major models –
one based on action at a distance and the other on a theory of fields
– resulted in a crucial advance of the electromagnetic theory
\cite{whittaker1989history}. Several factors influenced this process
and the final result was the formation of the currently accepted,
modern concept of the electromagnetic field. The first factor that
affected the evolution of the conceptualisation of electromagnetism
was the contribution of Michael Faraday, who conceived a model of
fields where interactions are transmitted as disturbances along the
medium, as opposed to the previous ideas of instantaneous transmission
at a distance \cite{harman1982energy}. This vision provided a
reasonable interpretation for various electrical and magnetic
interactions.

In mid-19th century, Maxwell, inspired by the ideas of Faraday and
William Thomson, developed a theory of the electromagnetic field based
on the idea of continually transmitting electrical and magnetic
actions. An essential component were the lines of force (introduced by
Faraday in his model) as states of mechanic ether governed by Newton’s
laws \cite{berkson2014fields}. In 1855, Maxwell published \textit{“On
  Faraday's Lines of Force”}, where he mathematically formulates the
lines of force based on the use of analogies with the movement of an
incompressible fluid \cite{darrigol2003electrodynamics}. An important
landmark in the development of the theory is the conceptualization of
ether as a quasi-material element that supports the lines of
force. Not surprisingly, in the following years Maxwell’s objective
was to find a mechanical model for ether that could describe the
electromagnetic field and allow to determine the speed of propagation
of interactions. This process is completed in 1861, with the
publication of \textit{“On Physical Lines of Force”} where he takes
his analogies to a different level by introducing an extremely complex
mechanical model of ether that allows him to develop an
electromagnetic theory of light and deduce the propagation speed of
electromagnetic waves \cite{nersessian2012faraday}.

One of the most original aspects of Maxwell’s theory was the
introduction of electrical displacement and the current of
displacement as another source of fields. In 1864, Maxwell clearly
described the meaning of both magnitudes.
\begin{quote}
    \textit{“Electrical displacement consists in the opposite
      electrification of the sides of a molecule or particle of a body
      which may or may not be accompanied with transmission through
      the body […] The variations of the electrical displacement must
      be added to the currents p, q, r to get the total motion of
      electricity…”} \cite[p~554]{maxwell1890scientific}
\end{quote}
Years later, in \textit{“A Treatise on Electricity and Magnetism”}, he
makes clear his position on the displacement current as a source of
magnetic fields:
\begin{quote}
\textit{“The current produces magnetic phenomena in its neighbourhood
  […] We have reason for believing that even when there is no proper
  conduction, but merely a variation of electric displacement, as in
  the glass of a Leyden jar during charge or discharge, the magnetic
  effect of the electric movement is precisely the same.”}
\cite[pp~144-145]{maxwell1873treatise}
\end{quote}
Given Maxwell's conception of space, and more specifically his
conviction about the existence of ether, the displacement current,
conceived as a consequence of the variation of the electric
displacement in any mechanical medium, was always associated with a
motion of bound charges.

The second important factor that influenced the development of the
electromagnetic theory was the progressive rejection of the ether
theory and the beginning of the adoption of an analytic
interpretation. Maxwell, aware of the limitations and difficulties
associated with his mechanical model of ether, decided to make it
independent of the electromagnetic field. In 1864, he published
\textit{“A Dynamical Theory of the Electromagnetic Field”} where he
presented eight equations of the electromagnetic field and an
electromagnetic theory of light that can be experimentally contrasted
\cite{harman1982energy}. In this task, he faced the difficulty of how
to interpret the set of developed equations. This is when the
analytical interpretation arises: the charge and the currents become
fundamental magnitudes and the field acts directly on the matter in
the point of interest, and as a consequence, the mechanisms of
continuous action imagined by Faraday disappear
\cite{berkson2014fields}. Nine years later, Maxwell published his most
important work, \textit{“A Treatise on Electricity and Magnetism”},
where he presented the whole electromagnetic theory in detail. It
continues with his analytical interpretation of the model, although he
still firmly believed in the existence of some mechanism subordinated
to Newton's laws \cite{berkson2014fields}.

Despite Maxwell’s formidable advances, there were still many things
that needed to be explained. These were, among others, the development
of an adequate theory of charge, the explanation of the properties of
the dielectrics, and the complete determination of the fields around
time-varying charges and currents. In addition, experiments needed to
be conducted to provide evidence which would support his field theory
\cite{whittaker1989history}. Years after Maxwell's death in 1887,
Heinrich Hertz experimentally demonstrated the existence of
electromagnetic waves, thus confirming the field theory as an
alternative to the theory of action at a distance
\cite{harman1982energy}. At the same time, Oliver Heaviside, used
vector calculus to analytically express the equations of Maxwell's
fields, writing the equations as we know them today
\cite{berkson2014fields}.

In 1892, Hendrik Lorentz took one more step in the development of the
classical electromagnetic field theory by assuming that all charged
bodies have charged particles and that ether is immobile and
unperturbed by their movement. This, consequently, means that the only
possible interpretation of the field equations is the analytical one
\cite{berkson2014fields}. Lorentz further assumed that electric and
magnetic fields are qualitatively different, although they are
produced and propagated outward by the charged particles
\cite{Roche_1987}. In contrast to Maxwell's view, Lorentz considered
an immobile ether and allowed for the existence of a displacement
current in the absence of matter, which would make it a simple term
directly proportional to the rate of change of the electric field
\cite{Roche_1998}.

The third important factor in the development of field theory were the
contributions that identified electric and magnetic fields as related
entities, the definitive rejection of the existence of ether and the
clarification of the sources of the fields. In 1905, Albert Einstein
postulated the theory of special relativity. In this framework, he
shows that Maxwell's laws adopt the same expressions in all inertial
reference frames and that the components of the electric and magnetic
fields relate to each other in the various inertial reference frames
by means of Lorentz’s transformations. He concluded that, contrary to
Maxwell's and Lorentz's theories, electric and magnetic fields should
not be considered as separate entities, but are in fact both part of
one entity: the electromagnetic field \cite{nersessian2012faraday}. In
this way, Einstein managed to resolve the asymmetries that appeared
when applying Maxwell's electrodynamics to moving bodies
\cite{einstein1905electrodynamics}. If an observer detects only a
magnetic field in a reference frame, another observer in motion will
always measure an electric and a magnetic field. Similarly, if an
electric field is measured in a reference frame, an electric and a
magnetic field will always be detected in another. From Einstein's
theory of special relativity onwards, the existence of ether lost its
\textit{raison d'être}, and by the 1920s its existence in the
scientific community was a thing of the past, and the analytic
interpretation of the field equations became the dominant one.

Another important milestone came in the 1960s when Oleg Jefimenko
elucidated the problem of the sources of the electromagnetic field
\cite{jefimenko1966electricity}. In his work he presents general
solutions for the electric and magnetic fields at a given instant and
point in space as a function of charge and current distributions which
can be expressed as \cite{jefimenko1989electricity}
\begin{eqnarray}
\label{jefimenkoE}
\mathbf{E}=\frac{1}{4 \pi
  \varepsilon_{0}}\int\left(\frac{\rho\left(r^{\prime},
  t^{\prime}\right)}{\left|\mathbf{r}-\mathbf{r^{\prime}}\right|^{3}}+\frac{\dot{\rho}\left(r^{\prime},
  t^{\prime}\right)}{c\left|\mathbf{r}-\mathbf{r^{\prime}}\right|^{2}}\right)\left(\mathbf{r}-\mathbf{r^{\prime}}\right)
d v^{\prime}\nonumber
\end{eqnarray}
\begin{eqnarray}
-\frac{1}{4 \pi \varepsilon_{0}}\int \frac{\dot{\mathbf{J}}\left(r^{\prime}, t^{\prime}\right)}{c^{2} \mid \mathbf{r}-\mathbf{r^{\prime} \mid}} d v^{\prime}
\end{eqnarray}
\begin{eqnarray}
\label{jefimenkoB}
\mathbf{B}=\frac{\mu_{0}}{4 \pi}
\int\left(\frac{\mathbf{J}\left(r^{\prime},
  t^{\prime}\right)}{\left|\mathbf{r}-\mathbf{r^{\prime}}\right|^{3}}+\frac{\dot{\mathbf{J}}\left(r^{\prime},
  t^{\prime}\right)}{c\left|\mathbf{r}-\mathbf{r^{\prime}}\right|^{2}}\right)
\times\left(\mathbf{r}-\mathbf{r^{\prime}}\right) d v^{\prime}
\end{eqnarray}
where the fields \textbf{E} and \textbf{B} are evaluated at position
$\mathbf{r}$ at time instant \textit{t}, being ${\textbf{r}}^{\prime}$
the distance from the origin of coordinates to charge density $\rho$
and current density \textbf{J}, $c$ is the speed of light and
$t^{\prime}=t-\left|{\textbf{r}}-{\textbf{r}^{\prime}}\right| / c$.

From these equations, Jefimenko concludes that the sources of the
fields are charge and current distributions
\cite{french2000maxwell,jefimenko2004presenting,heald2012classical,griffiths2013introduction,rosser2013interpretation,Tran_2018}. Given
that electric and magnetic fields have the same sources, when charge
and current distributions vary over time, both fields are generated
simultaneously and correlated through Faraday's and Ampere-Maxwell
laws \cite{Tran_2018}. All terms of Maxwell's laws in differential
form are evaluated at the same instant of time and at the same point
in space, therefore, they do not imply cause-and-effect relationships
and none of them can be the source of the other
\cite{hill2010rephrasing,hill2011reanalyzing}. From the point of view
of the relativistic formulation of electromagnetism, electric and
magnetic fields form a single entity, the electromagnetic field tensor
\cite{feynman1963feynman}. Just as it does not make sense to think
that one component of a vector is the cause of another component of
the same vector, it does not make sense that one component of the
tensor is the cause of another component of the same tensor
\cite{jefimenko2004presenting}. Hence, it is impossible to obtain the
temporal evolution of one of the fields knowing the temporal
evaluation of the other, since it ignores that they are components of
the same entity and, therefore, cannot interact with each other.

In summary, the historical and epistemological development of
electromagnetic field theory reveals important milestones in the
process: a) a mechanistic interpretation of electromagnetic
interaction (Faraday-Maxwell); b) a further development that lead to
an analytical interpretation based on Maxwell's laws which contradicts
the earlier model (Lorentz, Heaviside) and succeeds in experimentally
contrasting its prediction of electromagnetic waves (Hertz); c) an
evolution towards a unified theory of the electromagnetic field, with
Lorentz’s transforms (Einstein) and the definition of charge and
current distributions as the only sources of the electromagnetic field
(Jefimenko). Through this gradual elaboration of the theory with
Maxwell's laws as a framework which helped us analyse electromagnetic
interactions, certain fundamental epistemic features arise that allow
us to define a number of key concepts to be accounted for in the
textbooks, from an epistemologically grounded approach to
teaching. Next, we define these key ideas in relation to the classical
electromagnetic field theory, the equations that govern it and the
sources of the fields for university-level introductory physics:

\begin{itemize}
    \item KC1. The interpretation of Maxwell's laws implies non-causal
      relationships between the fundamental magnitudes of
      electromagnetic theory (\textbf{E}, \textbf{B}, \textbf{J} and
      $\rho$) at the same instant of time.
\end{itemize}
\begin{itemize}
    \item KC2. The sources of electromagnetic fields are charge and
      current distributions.
\end{itemize}
\begin{itemize}
    \item KC3. The cause of the fields in an electromagnetic wave (as
      well as that of electric and magnetic fields) at a certain
      instant is the distribution of charges and currents (generally
      time-varying) at an earlier time.
\end{itemize}
\begin{itemize}
    \item KC4. The electric and magnetic fields associate and form one
      single entity, the electromagnetic field, whose components are
      linked in different inertial systems by means of Lorentz’s
      transformations.
\end{itemize}

\section{\label{Meth}METHODOLOGY} 

\subsection{\label{Sel}Selection of textbooks \textcolor{black}{and assumptions of the analysis}}

In the sample of selected textbooks we use the most appropriate
non-probabilistic strategy, known as “purposive sampling” according to
Cohen et al \cite{cohen2007research}. Our aim is for the sample to be
representative, i.e. to reflect a situation common to the research
objective. The review took into account the bibliography recommended
in the introductory physics programmes presented on the websites of
Physics departments at prestigious universities in Spain, Latin
America and the USA. We chose 12 textbooks that introductory physics
academic programmes at universities recommend the most, with the
condition that they had been published in the last 15 years. In
principle, this condition makes it possible to incorporate the
abundant research in PER with regards to electromagnetism and
Maxwell's laws. Some textbooks that were very influential in
introductory physics programs in the past are not within the scope of
our current research and were left out. \textcolor{black}{In each
  textbook we analyse the chapters dealing with magnetic fields,
  Faraday's and Ampere-Maxwell's Laws, AC circuits, as well as with
  electromagnetic waves and special relativity. In Table \ref{codes}
  we show the name of each textbook of the sample and the chapters
  analysed.}

\begin{table*}
\textcolor{black}{
\begin{tabular}{|p{5.3cm}|p{5.5cm}|p{1.cm}|p{2.4cm}|}
\hline
\textit{Authors}&\textit{Titles}&\textit{Year}&\textit{Chap. analysed}\\ \hline
P. A. Tipler and G. Mosca&Physics for Scientists and Engineers&2008&26
to 30 and 39\\ \hline J. Walker, R. Resnick and
D. Halliday&Fundamentals of physics&2014&28 to 33 and 37\\ \hline
D. C. Giancoli&Physics for scientists and engineers with modern
physics&2014&27 to 31 and 36\\ \hline W. Bauer and
G. Westfall&University physics with modern physics&2014&27 to 31 and
35\\ \hline R. W. Chabay and B. A. Sherwood&Matter and
Interactions&2015&17 and 20 to 23\\ \hline D. M. Katz&Physics for
Scientists and Engineers: Foundations and Connections&2015&30 to 34
and 39\\ \hline T. A. Moore&Six ideas that shaped physics: Electric
and Magnetic Fields are Unified&2017&8 to 17 and appendix A\\ \hline
R. A. Serway and J. W. Jewett&Physics for scientists and
engineers&2019&28 to 33 and 38\\ \hline R. L. Hawkes, J. Iqbal,
F. Mansour, M. Milner-Bolotin and P. J. Williams&Physics for
scientists and engineers: an interactive approach&2019&24 to 27 and
30\\ \hline H. D. Young and R. A. Freedman&University Physics with
Modern Physics&2020&27 to 32 and 37\\ \hline R. Wolfson&Essential
University Physics&2020&26 to 29 and 33\\ \hline R. Knight&Physics for
Scientists and Engineers: strategic approach with modern
physics&2022&29 to 32 and 36\\ \hline
\end{tabular}
  \caption{Textbooks and chapters analysed.\label{codes}}}
\end{table*}

\textcolor{black}{The analysis was carried out by the four authors of
  the article who are professors or assistant professors in physics
  departments. They have extensive experience in teaching introductory
  physics courses, as well as courses in electrodynamics,
  thermodynamics and computational physics.}

\textcolor{black}{We have taken into account that the epistemological
  criteria defining the key concepts involve abstract concepts whose
  mathematical developments, such as the Jefimenko equations, are
  dealt with in electrodynamics courses. However, in the analysis we
  have restricted ourselves to the introductory level of the textbooks
  analysed.}  \textcolor{black}{In general, the approaches used are
  based on qualitative reasoning inspired on experimental facts rather
  than on mathematical formalism typical of advanced courses. So, we
  have analysed the key ideas as deeply as the textbooks themselves
  define them.}

\textcolor{black}{We designed a textbook-review protocol based on three
  criteria from the epistemological analysis and the key ideas defined
  in the previous section. We consider the theoretical explanations as
  well as the questions and solved examples used to illustrate the
  explanations. We have included the solved examples in the analysis
  because they illustrate the explanations of the theory and have
  allowed us to clarify the meaning that the authors explained in the
  theory sections. For the same reason, the analysis does not include
  the questions and exercises proposed at the end of chapters. Each
  criterion is developed through a protocol that constitutes a tool
  for analysis, which is explained in the following section.}

\subsection{\label{Ins}Instrument}

\textcolor{black}{We describe in this section the instruments of
  analysis for each criterium.}

\textit{Criterion 1. The treatment of Faraday and Ampère-Maxwell's
  laws in relation to the sources of the electric and magnetic field.}

Maxwell's laws do not imply cause-and-effect relationships (KC1) but
present mathematical relationships between different magnitudes at the
same instant of time. There cannot be a causal relationship between
the different terms of the equations. We understand causality
according to the principle of delayed action:

\textit{“there is always a time delay between the cause and the
  effect, the former being prior in time to the latter, so that
  (relatively to a given physical system, such as a reference system),
  C and E cannot be both distant in space and simultaneous.”}
\cite[p~62]{bunge2009causality}

Thus, charge and current distribution are the causes of the fields
which determine line integrals based on Faraday and Ampère-Maxwell
laws (KC2).

$$\textit{Faraday's Law in integral form }$$
\begin{eqnarray}
\label{intformfaraday}
\varepsilon=\oint \mathbf{E} \cdot \mathbf{d l}=-\frac{d}{d t} \int \mathbf{B} \cdot \mathbf{d A}
\end{eqnarray}

$$\textit{Ampère-Maxwell's law in integral form}$$ 
\begin{eqnarray}
\label{maxwell}
\oint \mathbf{B} \cdot \mathbf{d l}=\mu_{0} I+\mu_{0} \varepsilon_{0} \frac{d}{d t} \int \mathbf{E} \cdot \mathbf{d A}
\end{eqnarray}

Using Criterion 1 we will examine whether the following aspects are explained in the textbooks:

C1.1	Faraday and Ampère-Maxwell’s laws describe mathematical relationships between different magnitudes at the same instant of time.

C1.2	It is explained that the fields described in Faraday and Ampère-Maxwell’s laws do not imply causal relationships between electric or magnetic field operators and the generation of electric or magnetic field.

\textit{Criterion 2. Presenting the generation and propagation of electromagnetic waves in relation to their sources.}

Changes in the characteristics of an electromagnetic wave’s fields at
a certain point are associated with changes in the distribution of
charges or currents that generated it at an earlier time (KC3). This
aspect of the electromagnetic theory is key to understanding that
information propagates with finite speed and there exists a causal
relationship between the wave and its sources. Therefore, this
Criterion will examine whether the following aspects are explained in
the textbooks:

C2.1 Time-varying charge or current distributions over time generate
electromagnetic waves, linking the electromagnetic wave to its
sources.

C2.2 Electric and magnetic fields in electromagnetic waves are
associated with each other and are simultaneous in time without
establishing a causal relationship between them.

\textit{Criterion 3. Presenting the electromagnetic field as a single entity in relation to different reference systems.}

The analysis of the electromagnetic field in different reference
systems related by Lorentz transformations allows us to explain
clearly that there is one single electromagnetic field that presents
different components, in terms of \textbf{E} and \textbf{B}, to
different observers (KC4). In this respect, this Criterion will
analyse whether the following aspects are included in the textbooks:

C3.1 It is explained that, if an electric or magnetic field is
observed in one inertial reference frame, both fields are detected in
any other reference frame.

C3.2 Lorentz transformations for electric and magnetic fields are
introduced, the electromagnetic field is presented as a single entity
with two components: the electric field and the magnetic field.

\subsection{\label{Data}Data analysis}

In the first phase of data analysis, one of the researchers reviewed
six books from the sample according to the initial protocol that
included the three criteria. Subsequently, all four researchers met to
analyse and discuss the analysis protocol and suggest improvements on
the type of information expected. The analysis adopted does not
involve comparing a given textbook with an ideal book or ranking it in
relation to the criteria. In general, in each textbook we aim to
identify explanations, examples or questions which provide evidence
about the presentation of the topic in relation to the criteria
established, which we agreed to evaluate according to the following
classification:
 
\begin{itemize}
    \item Complies (C): the treatment of the theory, law or concept is considered appropriate to the Criterion set.
\end{itemize}
\begin{itemize}
    \item Partially complies (PC): some aspects of the Criterion are
      mentioned but not explained. When the content of a Criterion is
      not explicitly detected in the text but seems to be commented
      upon, it is included in this category.
\end{itemize}
\begin{itemize}
    \item Does not comply (NC): mentions elements that are not compatible with the theoretical framework or presents information that could lead to misunderstandings.
\end{itemize}
\begin{itemize}
    \item Does not address the issue (NT): does not address the issues
      included in the Criterion.
\end{itemize}
 
The revised protocol was applied by two of the researchers who
analysed the 12 textbooks independently. Subsequently, all researchers
met to discuss the results obtained and to clarify any doubts that
arose during the analysis process. We agreed that the four researchers
will carry out a second round of reviews, with the final protocol (see
Results below), to guarantee the validity of the analysis. The review
and validation protocols, as well as the two rounds of review in which
each textbook has been analysed by at least three researchers, allow
us to reduce possible bias in the information extraction of this
literature review. At the end of the second round, there was a strong
consensus among the researchers on whether or not the criteria were
met and on the comments on each textbook. \textcolor{black}{Finally, we
  determined Cohen's kappa coefficient, for the categories defined by
  the four evaluators. The statistic expresses a measure of the degree
  of agreement between evaluators who classify items into mutually
  exclusive categories and takes into account the possibility that
  they may agree by chance. We obtained an overall value of
  0.94. Values above 0.80 imply a high degree of agreement
  \cite{banerjee1999beyond}. In the next section we present the
  results of Cohen's kappa coefficient for each particular criterion.}

\section{\label{Res}RESULTS} 

\textcolor{black}{Here we describe the results obtained from the
  evaluation of each criterion. Summarised results for all criteria
  can be found in Table \ref{resumen}.}

\subsection{\label{Treatment}Textbooks treatment of Maxwell's laws in relation to electromagnetic field sources}

Criterion 1 assesses the ability to identify the sources of the fields
and to recognise that Maxwell's laws describe mathematical
relationships between different terms at the same instant in
time. \textcolor{black}{The Cohen's kappa coefficient obtained for this
  criterion was 1.0. Table \ref{criterio1} shows the results of the
  evaluation of the aspects included in criterion 1 for each of the
  textbooks.}

\begin{table*}\textcolor{black}{
\begin{tabular}{|c|c|c|}  
\hline
\textit{Authors}
 & \textit{Criterion 1.1} & \textit{Criterion 1.2}  \\ \hline
P. A. Tipler and G. Mosca&C&		NC  \\ \hline
J. Walker, R. Resnick and D. Halliday &C&		NC  \\ \hline
D. C. Giancoli
	&C&		NC  \\ \hline
W. Bauer and G. Westfall
	&C&		NC \\ \hline
R. W. Chabay and B. A. Sherwood
	&C&		C   \\ \hline
D. M. Katz
	&C&		NC  \\ \hline
T. A. Moore
	&C&		C  \\ \hline
R. A. Serway and J. W. Jewett
	&C&		NC  \\ \hline
R. L. Hawkes, J. Iqbal, \textit{et al.}  
	&C&		NC  \\ \hline
H. D. Young and R. A. Freedman
	&C&		NC  \\ \hline
R. Wolfson 
 &C&		NC  \\ \hline
R. Knight
 &C&		NC  \\ \hline
\end{tabular}
 \caption{\label{criterio1} Results corresponding to treatment of
   Faraday and Ampère-Maxwell's laws in relation to the sources of the
   electric and magnetic field. C1.1 refers to the explanation that
   these laws describe mathematical relationships between different
   magnitudes at the same instant of time while C1.2 implies that they
   do not imply causal relationships the fields.  In this table and
   the followings the keys are: C Complies, PC, Partially complies, NC
   does not comply, NT does not address.  }}
\end{table*}

As can be seen from Table \ref{criterio1}, all texts appropriately
explain that Faraday and Ampère-Maxwell’s laws describe mathematical
relationships between different magnitudes at the same instant of
time, giving correct analytical descriptions of the laws (Criterion
1.1). For example, Tipler and Mosca \cite[p~1034]{tipler2007physics}
state that:

\begin{quote}
\textit{“Faraday’s law states that the line integral of the electric
  field E around any closed curve C equals the negative of the rate of
  change of the flux of the magnetic field B through any surface S
  bounded by curve C.”}
\end{quote}
 
  Whereas Ampère-Maxwell law \cite[p~1034]{tipler2007physics}:
\begin{quote}
\textit{“...states that the line integral of the magnetic field B
  around any closed curve C equals 0 multiplied by the sum of the
  current I through any surface S bounded by the curve and the
  displacement current $I_{D}$ through the same surface.”}
\end{quote}
 
It is worth pointing out that the books analysed complement the
analytical descriptions of Maxwell's laws with different application
examples where it is clear that the different magnitudes involved are
evaluated at the same instant of time. For example, Bauer
\cite[p~880]{bauer2013university} presents the following problem of
Faraday's law:
 
\begin{quote}
\textit{“A current of $600mA$ is flowing in an ideal solenoid,
  resulting in a magnetic field of $0.025T$ inside the solenoid. Then
  the current increases with time, t, according to}
 $$i(t)=i_{0}\left[1+\left(2.4 s^{-2}\right) t^{2}\right]$$
\textit{ If a circular coil of radius $3.4cm$ with $N=200$ windings is located inside the solenoid with its normal vector parallel to the magnetic field (Figure 29.8), what is the induced potential difference in the coil at $t=2.0s$?"}
\end{quote}

In solving this problem, an expression is found for the magnetic field
generated by the solenoid as a function of the current
intensity. Then, to determine the induced emf, the magnetic field flux
is calculated and derived with respect to time, resulting in
\begin{eqnarray}
\varepsilon=-A B_{0}\left[2\left(2.4 s^{-2}\right) t\right]
\end{eqnarray}
where A is the cross-sectional area of the solenoid, while $B_{0}$ is its magnetic field in $t=0s$. Finally, the induced emf at $t=2.0s$ is found. The presentation of problems like the one described, where the emf has to be determined for a certain value of $t$, allows to infer that the terms involved are evaluated at the same instant of time.

According to Criterion 1.2, the fields described in Faraday and
Ampère-Maxwell’s laws do not imply causal relationships between
electric or magnetic field operators and the generation of electric or
magnetic field. This aspect of Maxwell's laws is only taken into
account in two of the textbooks analysed. One of the texts that
complies with the Criterion is Moore's \cite[p~271]{moore2017six}
where, explaining Faraday's law, he explicitly states that:

\begin{quote}
\textit{“I have been very careful to state that this is the electric
  field that is correlated with the changing magnetic field, not
  created by that field. Electromagnetic fields are created only by
  stationary or moving charged particles."}
\end{quote}

However, 10 of the 12 textbooks analysed interpret Faraday and
Ampère-Maxwell's laws as describing cause-and-effect relationships
between different terms, such that the time-varying magnetic field is
the source of the electric field (Faraday's law) and vice versa, the
time-varying electric field is the source of the magnetic field
(Ampère-Maxwell's law). This interpretation is explicitly presented,
for example, when summarising the physical meaning of Maxwell's
laws. It states:
\begin{quote}
\textit{“Faraday’s law: An electric field can also be created by a changing magnetic field.}

\textit{Ampère-Maxwell law, first half: Currents create a magnetic field.}

\textit{Ampère-Maxwell law, second half: A magnetic field can also be  created by a changing electric field."} \cite[p~939]{knight2022physics}
\end{quote}

Interpretations like this are reinforced through the analysis of
different situations where Faraday and Ampère-Maxwell’s laws are
applied. One of the most representative cases is the calculation of
the electric field around a solenoid where the current intensity
changes with time. For example, the following problem is entitled
\textit{“E produced by changing B”}:

\begin{quote}
\textit{“A magnetic field B between the pole faces of an electromagnet
  is nearly uniform at any instant over a circular area of radius as
  shown in Figs. 27a and b. The current in the windings of the
  electromagnet is increasing in time so that changes in time at a
  constant rate $d{{B}}/dt$ a\textit{t each point. Beyond the circular
    region} $\left(r>r_{0}\right)$ \textit{we assume $B=0$ at all
    times. Determine the electric field} E at any point P a distance r
  from the center of the circular area due to the changing B."}
\cite[p~892]{giancoli2014physics}
\end{quote}

In this problem, the time-varying magnetic field is explicitly
described as the cause of the electric field, however, both fields are
generated at the same time and have a common cause: the varying
current.

In the case of Ampère-Maxwell law, the 10 textbooks that do not comply
with Criterion 1.2 provide explanations of the effects of the
displacement current that could be a source of misunderstandings, for
example:

\begin{quote}
\textit{“Although changing electric flux isn’t the same thing as
  electric current, it has the same effect in producing a magnetic
  field. For this reason Maxwell called the term} $\varepsilon_{0} d
\Phi_{E}/dt$ \textit{the displacement current. The word displacement
  has historical roots that don’t provide much physical insight. But
  current is meaningful because displacement current is
  indistinguishable from real current in producing magnetic fields."}
\cite[p~582]{wolfson2020essential}
\end{quote}

Assuming that the displacement current generates a magnetic field may
lead students to believe that there is a cause-and-effect relationship
between the different terms of Ampère-Maxwell law. Even more so, if we
take into account that one of the first applications of the
displacement current is the calculation of the magnetic field between
the plates of a capacitor that is charging. For example, in one of the
textbooks analysed, the following situation arises:

\begin{quote}
\textit{“You might well ask at this point whether displacement current
  has any real physical significance or whether it is just a ruse to
  satisfy Ampère’s law and Kirchhoff’s junction rule. Here’s a
  fundamental experiment that helps to answer that question. We take a
  plane circular area between the capacitor plates (Fig. 29.23). If
  displacement current really plays the role in Ampère’s law that we
  have claimed, then there ought to be a magnetic field in the region
  between the plates while the capacitor is charging."}
\cite[p~972]{young2019university}
\end{quote}

Next, the magnetic field between the capacitor plates is determined
from the Ampère-Maxwell law. To do this, a closed curve placed between
the capacitor plates is taken and the displacement current through the
surface delimited by this curve is determined. Then, the displacement
current is expressed as a function of the conduction current
\textit{I}, and the magnetic field is determined at a distance
\textit{r} from the symmetry axis of the capacitor, for points such
that $r \leq R$, being \textit{R} the radius of the plates, we obtain:
\begin{eqnarray}
\label{Bcap}
B=\frac{\mu_{0}I}{2\pi}\frac{r}{R^{2}}
\end{eqnarray}

In arriving at equation \ref{Bcap}, it is concluded that: 
\begin{quote}
\textit{“When we measure the magnetic field in this region, we find
  that it really is there and that it behaves just as Eq. (29.17)
  predicts. This confirms directly the role of displacement current as
  a source of magnetic field. It is now established beyond reasonable
  doubt that Maxwell’s displacement current, far from being just an
  artifice, is a fundamental fact of Nature."}
\cite[p~972]{young2019university}
\end{quote}

In contrast, in the textbooks that comply with the Criterion, they do
not associate the term $\varepsilon_{0} d \Phi_{E}/dt$ with a
current. For example, when analysing the different terms of
Ampere-Maxwell’s law, Chabay and Sherwood
\cite[p~941]{chabay2015matter} conclude that:

\begin{quote}
\textit{“We interpret the Ampere–Maxwell law as saying that a
  time-varying electric field is always associated with a magnetic
  field."}
\end{quote}

The fact that Ampere-Maxwell law can be used to calculate a magnetic
field whose circulation is directly proportional to the speed with
which the electric field flux varies does not mean that it is
generated by it. As discussed in Section \ref{Epistemology}, a
time-varying electric field does not generate a magnetic field; the
sources of the magnetic field between the plates of a charging
capacitor are the surface currents in its plates and the conduction
currents in the wires
\cite{Rosser1976,french2000maxwell,milsom2020untold,Hyodo_2022}.

The results indicate that, depending on the context, 10 textbooks
recognise the presence of four possible sources of fields \textbf{E}
and \textbf{B}: a) charges; b) current; c) variation of \textbf{E}
generates \textbf{B}; d) variation of \textbf{B} generates
non-coulombic \textbf{E}. They do not consider it contradictory to
indicate charges and currents as sources on the one hand, and
time-varying electric and magnetic fields on the other.

\subsection{\label{Sources}Presenting the generation and propagation of electromagnetic waves in relation to their sources}

The study of the generation and propagation of electromagnetic waves
provides a unique opportunity to clarify the sources of the fields, to
recognise that information does not propagate instantaneously and that
electric and magnetic fields in waves are correlated but there is no
causal relationship between them. These aspects are assessed through
Criterion 2. \textcolor{black}{The Cohen's kappa coefficient obtained
  for this criterion was 0.88. Table \ref{criterio2} shows the results
  of the evaluation of the aspects included in criterion 2 for each of
  the textbooks.}

\begin{table*}
\textcolor{black}{
\begin{tabular}{|c|c|c|}  
\hline
\textit{Authors}
 & \textit{Criterion 2.1} & \textit{Criterion 2.2}  \\ \hline
P. A. Tipler and G. Mosca
& C & C   \\ \hline
J. Walker, R. Resnick and D. Halliday 
& C & NC  \\ \hline
D. C. Giancoli
& C & NC  \\ \hline
W. Bauer and G. Westfall
& C & C   \\ \hline
R. W. Chabay and B. A. Sherwood
 &  C& C   \\ \hline
D. M. Katz
	& NC& NC  \\ \hline
T. A. Moore
	& C &  C  \\ \hline
R. A. Serway and J. W. Jewett
& C & NC  \\ \hline
R. L. Hawkes, J. Iqbal, \textit{et al.} 
	& C & NC  \\ \hline
H. D. Young and R. A. Freedman
& C & NC  \\ \hline
R. Wolfson 
 & C & NC  \\ \hline
R. Knight
 & NC &NC  \\ \hline
\end{tabular}
 \caption{\label{criterio2} Summary of the results about the
   generation and propagation of electromagnetic waves in relation to
   the sources of the electromagnetic waves.  C2.1 implies the
   explanation that time-varying charge or current distributions
   generate electromagnetic waves while C2.2 implies in
   electromagnetic waves both fields are associated with each other
   and are simultaneous in time without establishing a causal
   relationship between them.}  }
\end{table*}

Table \ref{criterio2} reveals that a large majority of the textbooks
analysed explicitly describe that the sources of the waves are the
accelerated charges. They do this using different approaches and
depths, often relying on the analysis of radiation emitted by
oscillating charges in an antenna.

Especially noteworthy is the textbook by Hawkes et
al. \cite{hawkes2019physics}, which takes a different approach based
on the analysis of the fields generated by a point charge in different
inertial reference frames. A point charge performing an URM generates
an electric field and a magnetic field around itself. An observer at
rest with respect to the charge detects only a static electric field,
so for such an observer the charge does not radiate electromagnetic
waves. With this in mind, Hawkes et
al. \cite[p~974]{hawkes2019physics} state:

\begin{quote}
\textit{“The laws of physics must be the same in all frames that move
  with a uniform velocity with respect to each other. So, if a charge
  cannot emit electromagnetic radiation in its own reference frame, it
  cannot emit electromagnetic radiation in any frame moving with
  uniform velocity with respect to the charge. Thus, a charge can emit
  electromagnetic radiation in free space only when accelerating."}
\end{quote}

Regardless of the approach used, we can see a general agreement on the
fact that the electromagnetic wave consists of a disturbance of the
propagating electromagnetic field and that the origins of such
disturbances are accelerated charges. For instance, Wolfson
\cite[p~593]{wolfson2020essential} writes:

\begin{quote}
\textit{“All it takes to produce an electromagnetic wave is a changing
  electric or magnetic field…Ultimately, changing fields of both types
  result when we alter the motion of electric charge. Therefore,
  accelerated charge is the source of electromagnetic waves."}
\end{quote}

However, when electromagnetic wave generation processes are explained,
we must pay attention to the language if we wish to avoid possible
misunderstandings. In this respect, two of the textbooks analysed do
not comply with Criterion 2.1. Depending on the context, they suggest
two possible sources of electromagnetic waves: accelerated charges or
a time-varying electric field. For example, when explaining the
generation of the waves, one of the textbooks describes the
transmitter of Hertz’s original experiment, which consists of a
voltage source connected to an induction coil using a switch and a
parallel connected capacitor made of two metal rods separated by a gap
and two spheres connected at their ends. It arrives to the following
conclusion:

\begin{quote}
\textit{“When the transmitter is turned on, a strong electric field
  builds up in its gap. The electric field ionizes the air, and
  acceleration of the resulting free electrons causes more ionization
  until the air conducts a spark. The plates of the capacitor charge
  and discharge periodically, while the sparks in the gap oscillate at
  the natural frequency of the LC circuit. The electric field
  oscillation creates an electromagnetic wave."}
\cite[p~1094]{katz2015physics}
\end{quote}

In our analysis of the way the textbooks address the relationship
between electric and magnetic fields in electromagnetic waves we found
that every case demonstrates that in a plane wave the fields are in
phase and are mathematically linked according to the equation:
\begin{eqnarray}
\label{E/B}
\frac{E(t)}{B(t)}=c
\end{eqnarray}

Which involves, according to Serway and Jewett \cite[p~881]{serway2018physics}:

\begin{quote}
\textit{“...at every instant, the ratio of the magnitude of the electric field to the magnitude of the magnetic field in an electromagnetic wave equals the speed of light."} 
\end{quote}

In contrast, when explaining the propagation of electromagnetic waves,
only one third of the selected textbooks explain propagation without
establishing causal relationships between electric and magnetic fields
(Criterion 2.2). For instance, Bauer \cite[p~943]{bauer2013university}
writes that:

\begin{quote}
\textit{“Electromagnetic waves consist of electric and magnetic
  fields, can travel through vacuum without any supporting medium, and
  do not involve moving charges or currents. The existence of
  electromagnetic waves was first demonstrated in 1888 by the German
  physicist Heinrich Hertz (1857–1894). Hertz used an RLC circuit that
  induced a current in an inductor that drove a spark gap. A spark gap
  consists of two electrodes that, when a potential difference is
  applied across them, produce a spark by exciting the gas between the
  electrodes. Hertz placed a loop and a small spark gap several meters
  apart. He observed that sparks were induced in the remote loop in a
  pattern that correlated with the electromagnetic oscillations in the
  primary RLC circuit. Thus, electromagnetic waves were able to travel
  through space without any medium to support them."}
\end{quote}

We note that the author describes the propagation of waves in a way
that is simple and understandable for students without referring to
any interaction between the electric and magnetic fields of the
electromagnetic wave.

In line with this, Chabay and Sherwood \cite[p~947]{chabay2015matter}
state that:
\begin{quote}
\textit{ “Maxwell concluded that light must be a combination of time-varying electric and magnetic fields that can propagate through otherwise empty space, far from any charges or currents."} 
\end{quote}

Later, when referring to the electric and magnetic fields of the electromagnetic wave, Chabay and Sherwood \cite[p~948]{chabay2015matter} write that:

\begin{quote}
\textit{“...a pulse of radiation must contain both electric and
  magnetic fields, so that the pulse can move forever through
  otherwise empty space, far from any charges. The transverse electric
  field is accompanied by a transverse magnetic field perpendicular to
  the electric field, with} $\mathbf{E} \times \mathbf{B}$ \textit{in
  the direction of propagation."}
\end{quote}

It can be deduced from the explanation that the electric and magnetic
fields in electromagnetic waves do not interact with each other, being
generated at the same time. The explanation given by Chabay and
Sherwood is in accordance with equation \ref{E/B}, according to which
the fields in a point in space always vary in the same way and
simultaneously, so that one field cannot be the cause of the other.

However, two thirds of the textbooks analysed consider that the
propagation of electromagnetic waves lies in a supposedly
self-sustaining mechanism, originated by a time-varying electric field
that is supposed to generate a time-varying magnetic field that would
in turn generate an electric field and so on. Let us consider the
following examples:

\begin{quote}
\textit{“Once an electromagnetic wave is generated, it propagates on
  its own and does not require any physical medium to continue its
  propagation. The wave generates itself because a time-varying
  electric field generates a time-varying magnetic field, which in
  turn generates a time-varying electric field. This process is
  continually repeated."} \cite[p~962]{hawkes2019physics}
\end{quote}

\begin{quote}
\textit{“First consider the magnetic field. Because it varies
  sinusoidally, it induces (via Faraday’s law of induction) a
  perpendicular electric field that also varies sinusoidally. However,
  because that electric field is varying sinusoidally, it induces (via
  Maxwell’s law of induction) a perpendicular magnetic field that also
  varies sinusoidally. And so on. The two fields continuously create
  each other via induction, and the resulting sinusoidal variations in
  the fields travel as a wave—the electromagnetic wave."}
\cite[p~976]{walker2014halliday}
\end{quote}

\subsection{\label{electrofield}Presenting the electromagnetic field as a single entity in relation to different reference systems}

The special theory of relativity laid the foundation for a new
conception of electric and magnetic fields by no longer considering
them as separate entities but parts of a single entity, the
electromagnetic field, which adopts different expressions depending on
the reference system. These aspects of the electromagnetic field are
assessed through Criterion 3. \textcolor{black}{The Cohen's kappa
  coefficient obtained for this criterion was 0.94. Table
  \ref{criterio3} shows the results of the evaluation of the aspects
  included in criterion 3 for each of the textbooks.}

\begin{table*} 
\textcolor{black}{
\begin{tabular}{|c|c|c|}  
\hline
\textit{Authors}
 & \textit{Criterion 3.1} & \textit{Criterion 3.2}  \\ \hline
P. A. Tipler and G. Mosca
& NT & NT \\ \hline
J. Walker, R. Resnick and D. Halliday 
& NT & NT  \\ \hline
D. C. Giancoli
& NT & NT  \\ \hline
W. Bauer and G. Westfall
& NT & NT  \\ \hline
R. W. Chabay and B. A. Sherwood
& C  & C   \\ \hline
D. M. Katz
	& NT & NT  \\ \hline
T. A. Moore
& C  & C   \\ \hline
R. A. Serway and J. W. Jewett
& NT & NT  \\ \hline
R. L. Hawkes, J. Iqbal,  \textit{et al.} 
& C & NT   \\ \hline
H. D. Young and R. A. Freedman
& NT & NT  \\ \hline
R. Wolfson 
 & C  & NT  \\ \hline
R. Knight
 & C  & C   \\ \hline
\end{tabular}
 \caption{\label{criterio3} Results of the criterion 3 evaluating
   whether the textbooks selected present the electromagnetic field as
   a single entity.  C3.1 refers to, if an electric or magnetic field
   is observed in one inertial reference frame, both fields are
   detected in any other reference frame while C3.2 is related to the
   presentation of the Lorentz transformations and the electromagnetic
   field as a single entity.}  }
\end{table*}

Research in Physics Education has advocated the inclusion of a
simplified relativistic approach to electromagnetism in introductory
physics courses, in order to establish coherence and encourage the
construction of a synthesised view of electric and magnetic fields as
complementary facets of the same entity
\cite{galili1997changing}. However, more than half of the textbooks
analysed do not address any of the aspects contained in Criterion
3. As far as the analysis of electric and magnetic fields in different
reference systems is concerned, fewer than half of the textbooks deal
with the subject (Criterion 3.1), while field transformations are
dealt with by three of the textbooks analysed (Criterion 3.2).

The place within the study of electromagnetism where the aspects
mentioned in Criterion 3 are addressed changes depending on the text
analysed. For example, Chabay and Sherwood
\cite[p~678]{chabay2015matter}, following Biot-Savart’s law, explain a
“thought experiment” with fictitious students. Jack is sitting in the
classroom and has an electrically charged tape and a compass. Since,
in his opinion, the tape is at rest, it detects an electric field, but
the compass does not deflect. Meanwhile, Jill, who also has a compass
in her hands, runs at high speed in front of Jack. Jill detects an
electric field, but because in her opinion the charged tape is moving,
her compass deflects and detects a magnetic field. Thus, Chabay and
Sherwood \cite[p~678]{chabay2015matter} conclude:

\begin{quote}
\textit{“Up until now we have implied that electric fields and
  magnetic fields are fundamentally different, but this “thought
  experiment” shows that they are in fact closely related."}
\end{quote}

Furthermore, Wolfson \cite[p~682]{wolfson2020essential} deals with the
relative nature of the fields in the chapter on special relativity,
after having dealt with Maxwell's laws and electromagnetic waves. To
do so, he starts from the example of a charge performing an URM with
respect to a certain inertial reference frame and compares the fields
that would be detected by an observer located in that frame versus the
other who is at rest with respect to the charge. Thus, Wolfson
\cite[p~682]{wolfson2020essential} conclude:

\begin{quote}
\textit{“So electric and magnetic fields aren’t absolutes; what one
  observer sees as a purely electric field another may see as a mix of
  electric and magnetic fields, and vice versa. You can think of the
  electric and magnetic field as components of a more fundamental
  electromagnetic field; how that field breaks out into electric and
  magnetic fields depends on your frame of reference."}
\end{quote}

Within the textbooks that address field transformations and present
the electromagnetic field as a single entity, we come across two very
different approaches to the problem. Knight’s textbook
\cite{knight2022physics}, for example, has a similar structure to most
of the texts analysed in terms of the order and presentation of the
topics and devotes an entire chapter to discuss the electromagnetic
field, its transformations and the electromagnetic waves. He starts by
comparing the forces on a charged particle in a magnetic field, for
observers in different reference frames and assuming that the force is
invariant, he concludes \cite[p~930]{knight2022physics}:

\begin{quote}
\textit{“Whether a field is seen as “electric” or “magnetic” depends
  on the motion of the reference frame relative to the sources of the
  field."}
\end{quote}

Knight then deduces an approximation for the transformation of the
fields at low velocities and introduces the notion of the
electromagnetic field \cite[p~932]{knight2022physics}:

\begin{quote}
\textit{“We can no longer believe that electric and magnetic fields
  have a separate, independent existence. Changing from one reference
  frame to another mixes and rearranges the fields. Different
  experimenters watching an event will agree on the outcome, such as
  the deflection of a charged particle, but they will ascribe it to
  different combinations of fields. Our conclusion is that there is a
  single electromagnetic field that presents different faces, in terms
  of E and B, to different viewers."}
\end{quote}

Furthermore, in Moore's textbook \cite{moore2017six} the
electromagnetic field is introduced before Maxwell's own laws and
after having presented the fields \textbf{E} and \textbf{B} in
situations with constant charges and currents. Moore, after analysing
the electric and magnetic forces on an electron in the neighbourhood
of a conductor through which current flows for different inertial
reference systems and applying the principle of relativity and
Lorentz’s contraction, concludes \cite[p~187]{moore2017six}:

\begin{quote}
\textit{“...we must recognize that electric and magnetic fields are
  but two different aspects of an electromagnetic field, whose reality
  and effects are frame independent, but whose division into electric
  and magnetic parts is not. The degree to which we interpret a given
  electromagnetic field as being electric and/or magnetic depends on
  one’s choice of reference frame. For example, the electromagnetic
  field of a charged particle at rest is purely electric, but is a
  mixture of electric and magnetic fields in a frame where the
  particle is moving."}
\end{quote}

Moore \cite{moore2017six} then introduces Lorentz transformations for
fields and applies them in order to study the electromagnetic field
generated by a charged particle.

\section{\label{Disc}DISCUSSION}

Within the framework of the defined criteria, in this section we
discuss the results of the research questions that examined how the
selected textbooks introduced and explained the nature of
electromagnetic fields. We start by focusing on the descriptions of
Maxwell’s equations, then we analyse how the textbooks explain the
generation and propagation of electromagnetic waves. We conclude by
looking at their explanations as to how the electromagnetic field is
presented depending on the frame of reference chosen.

Regarding the introduction of Maxwell’s equations and sources of
magnetic fields, we found that all the textbooks gave clear
explanations of the mathematical relationships between the various
terms in Faraday’s and Ampère–Maxwell’s laws, but most only offered
causal interpretations, identifying four possible sources of
electromagnetic fields (Criteria 1.1 and 1.2, Table
\ref{criterio1}). However, an epistemological analysis of the nature
of electromagnetic fields reveals that the field sources are charge
and current distributions. The sometimes confusing presentation of
Maxwell’s equations may play a part in some of the difficulties
students display in their understanding of electric and magnetic
fields, for instance, when they mix up force with field or do not know
how to define the main characteristics of an electric or magnetic
field
\cite{guisasola2004difficulties,zuza2018introductory,campos2019}. If
the sources of electromagnetic fields were introduced in line with the
epistemological analysis, it would promote easier comprehension and a
better understanding of electric and magnetic fields, and it might
help prevent students from developing a cause-and-effect
interpretation of Faraday’s and Ampère–Maxwell’s laws. In the case of
Faraday’s law, for example, a clear account of how a changing current
in a solenoid generates both an electric field and a magnetic field
(caused by the current) would allow students to visualise how the law
describes the mathematical relationship between the circulation of an
electric field around a closed curve and the magnetic field flux
variation; and without implying a cause-and-effect relationship
between the two terms.

As for Ampère–Maxwell’s law, we found that most textbooks in the
sample (Criterion 1.2) attribute the cause of the magnetic field
between the plates of a parallel plate capacitor to the displacement
current. This interpretation has endured over time and is a possible
cause of misunderstandings. Nevertheless, the epistemological analysis
of the classical theory of electromagnetism shows that the magnetic
field is caused by the conduction currents in the capacitor plates. To
avoid confusion, Rosser \cite{Rosser1976} suggested renaming
$\varepsilon_{0} d \Phi_{E}/dt$ the “Maxwell term” and discontinuing
the use of “displacement current”. Incomplete explanations may
complicate the students’ ability to understand Maxwell’s equations and
their application to concrete phenomena by confusing flux changes with
field sources
\cite{zuza2012rethinking,Kuo,suarez_marti_zuza_guisasola_2022,suarez2022unified}.

When we analysed how the textbooks presented the generation and
propagation of electromagnetic waves in relation to their sources, we
found that most gave suitable explanations about the origin of waves,
i.e., that they are caused by time-varying charge or current
distributions. Furthermore, two-thirds of the textbooks used arguments
based on a supposed mechanism of mutual generation between the
time-varying fields to account for the propagation of electromagnetic
waves (Criteria 2.1 and 2.2, Table \ref{criterio2}). This kind of
explanation may have arisen from a causal interpretation of the
differential equations in which time was considered an independent
variable \cite{bunge2009causality}. Thus, if we apply the differential
form of Faraday’s law to the fields in an electromagnetic wave, it is
reasonable to conclude that the time-varying magnetic field is the
cause of the electric field. However, the differential equations do
not imply that the change in one phenomenon over time is the cause of
the other; they only help affirm, in the case of Faraday’s law, for
example, that the rate of change of the magnetic field is associated
with a spatial variation in the electric field. The causal
interpretation between different phenomena connected by a law is based
on the explanations of the phenomena and described with semantic rules
of correspondence \cite{bunge2009causality}, as is the case with
Newton’s second law, where the differential equation can be
interpreted as a typical lineal causal relationship in which the force
is the cause of the acceleration. The sort of incomplete descriptions
of electromagnetic waves that we observed in some of the textbooks
could culminate in students finding it hard to assimilate the concepts
behind electromagnetic waves and their propagation
\cite{Ambrose1999,Podolefsky2007}.

We have shown that more than half of the textbooks in the study did
not include Einstein’s contributions to the development of
electromagnetic theory (Criteria 3.1 and 3.2, Table
\ref{criterio3}). Yet, an assessment of the asymmetries that appear in
electromagnetic phenomena when Maxwell’s equations are applied within
different inertial frames of reference could help students better
understand the relativistic nature of electromagnetic fields and the
theory’s coherence. Since electromagnetic induction is explained from
the perspective of different observers, for example, then it provides
the perfect opportunity to explain away these supposed asymmetries in
the electromagnetic field and encourage students to think about the
explanatory power of electromagnetic theory
\cite{galili1997changing,potters2017studying,chabay2015matter,knight2022physics}

\section{\label{Concl}CONCLUSIONS} 

Scientific theories are characterised by their consistency and
predictive power. These qualities are broadly accepted, but it is
often hard to transmit them to physics students. The analysis of
textbooks is justified by the importance of its influence in both
teaching and learning in the classrooms where sciences are taught. In
this paper, we research into the way the nature of the electromagnetic
field is presented in a dozen of textbooks chosen amongst those most
frequently used to teach Introductory physics at university.

Our analysis, which focused on the epistemological development of
electromagnetic theory, highlights that many of the textbooks
exhibited inconsistencies in their approach to electromagnetic fields
and their relationship with their sources, Maxwell’s equations and
electromagnetic waves. \textcolor{black} {In most of the textbooks
  analysed there is a tendency to make cause-effect interpretations
  between the different terms of Maxwell's equations. Ten of the
  twelve books analysed (see Table \ref{criterio1}) explicitly state
  that Faraday's law implies that a varying magnetic field generates
  an electric field and that the Ampère-Maxwell law implies that a
  varying electric field generates a magnetic field, without
  discussing other more intuitive sources such as electric charges at
  rest or in motion previously discussed. Therefore, one could
  interpret that there are multiple sources of the electromagnetic
  field, including theoretical constructs such as the flux of the time
  varying electric field. However, as we have contrasted in the
  epistemological analysis, the only sources of any electromagnetic
  field are charge and current distributions. See, for example, how
  this is explained at the introductory physics level in one of the
  books reviewed \cite[p~271]{moore2017six}.}

\textcolor{black}{Regarding the propagation of electromagnetic waves,
  eight of the twelve textbooks analysed explain it through a
  mechanism of mutual induction, where a variable electric field
  generates a variable magnetic field that in turn generates a
  variable electric field and so on (see Table \ref{criterio2}). In
  this way the fields continuously create each other and travel
  generating the electromagnetic wave. This explanation may lead
  students to think that there are multiple sources of the
  electromagnetic field and promote an incomplete view by considering
  the electric field and the magnetic field as two independent
  entities. This ignores the fact that these fields have a common
  cause and are components of the same entity, the electromagnetic
  field, and therefore cannot interact with each other
  \cite{jefimenko2004presenting}.}

The inconsistencies detected in our analysis suggest that
electromagnetism should be taught with an emphasis on the relationship
between the different terms in Maxwell’s equations so that students
can appreciate that both the electric and magnetic fields in an
electromagnetic wave must simultaneously fulfil Maxwell’s
equations. As such, students couldn’t possibly imagine one field
existing without the other and it would be meaningless to ask which is
the cause and which is the effect. Each field depends on the other and
neither can be considered the source of the other, hence there is a
relationship of circular causality between the fields
\cite{halbwachs1971reflexions}. The fields of an electromagnetic wave
constitute a single entity: the electromagnetic field, which is
produced by accelerated charges and propagates at the speed of
light. We believe that if there were a greater emphasis on these
points, along with a suitable description of electromagnetic field
sources, students would be able to develop a more accurate mental
construct of the generation and propagation of electromagnetic waves
by simplifying and clearly separating the two processes. This would
set the stage to address the concept of the electromagnetic field and
connect both processes through the notion that the fields propagate at
a finite velocity.

\textcolor{black}{In respect of the criterion 3, considering the
  electromagnetic field as a single entity in relation to different
  reference systems, we find that, unlike the other criteria analysed,
  less than half of the books address the subject (see Table
  \ref{criterio3}). The books that mention this aspect do so in line
  with the epistemological development of the theory. However, it
  would be desirable for introductory physics textbooks to
  qualitatively analyze these aspects of the electromagnetic field, as
  it is a key topic for establishing coherence in the theory and
  providing a unified view of the field.}

An introduction to the Lorentz transformations can be used to resolve
the asymmetries and contradictions that arise when electromagnetic
theory is applied to different inertial frames of reference and to
formally describe the singularity of electromagnetic
fields. Considering that the special theory of relativity is covered
in more advanced courses, it is hard to imagine the inclusion of the
Lorentz transformations in introductory courses on
electromagnetism. Nevertheless, we believe it is feasible to introduce
a simplified relativistic framework for electromagnetism that is
consistent with courses on classical mechanics. The invariance of the
force could be used to resolve the contradictions in the predictions
made for different observers, which would bestow the Lorentz force a
more fundamental status compared to that of the electric and magnetic
force \cite{Preti_2009,Heras2010}. This approach could provide a
suitable framework for students to understand that the field
transformations originate from the need for simple physical
situations, by using links between the perspectives of two observers
in relative motion to each other
\cite{galili1997changing,knight2022physics}. Although Galilean
transformations do not resolve the asymmetries and contradictions that
arise when Maxwell’s equations are applied to different inertial
frames of reference, their introduction means students can appreciate
the inherently relativistic nature of electromagnetism and develop a
consistent concept of the electromagnetic field in keeping with
current scientific knowledge.

\textcolor{black}{The inconsistencies in the presentation of the
  electromagnetic field found in the textbooks in the sample coincide
  with some of the learning difficulties detected by research in PER
  \cite{Ambrose1999,guisasola2004difficulties,guisasola2008gauss}. This
  leads us to think that similar attention to PER results devoted by
  textbook authors and curriculum designers in other physics topics
  (e.g., in mechanics) would be desirable in the topic of
  electromagnetic field sources and their propagation. This attention
  could help students to a greater understanding. In the table
  \ref{resumen}, we note that two books stand out that meet all the
  assessed aspects, Matter and interactions \cite{chabay2015matter}
  and Six ideas that shaped physics \cite{moore2017six}. These books
  are distinctive in that they pay particular attention to PER
  outcomes, and because their use in the classroom has been subject to
  evaluation in relation to the learning achieved by students
  \cite{Kohlmyer,mooreinstructor}.This is not to say that just paying
  attention to PER results when explaining theory would improve
  student learning, but rather that taking these research findings
  into account in textbooks is a factor that can improve learning.} We
hope this study provides some ideas to inspire the development of new
teaching materials that are more in line with scientific practice and
progress and the current status of classical electromagnetic theory.

\begin{table*}
\textcolor{black}{
\begin{tabular}{|p{6cm}|c|c|c|c|c|c|c|}
\hline
\textit{Authors}  & \textit{Cr. 1.1} & \textit{Cr. 1.2} &\textit{Cr.  2.1} &\textit{Cr. 2.2}&\textit{Cr. 3.1 }&\textit{Cr. 3.2} & \textit{Criteria satisfied} \\ \hline
P. A. Tipler and G. Mosca
&C&		NC & C & C  & NT & NT & 3 \\ \hline
J. Walker, R. Resnick and D. Halliday 
&C&		NC & C & NC & NT & NT & 2 \\ \hline
D. C. Giancoli
&C&		NC & C & NC & NT & NT & 2 \\ \hline
W. Bauer and G. Westfall&C&		NC & C & C  & NT & NT & 3 \\ \hline
R. W. Chabay and B. A. Sherwood
	&C&		C  &  C& C  & C  & C  & 6 \\ \hline
D. M. Katz
	&C&		NC & NC& NC & NT & NT & 1 \\ \hline
T. A. Moore
	&C&		C & C &  C  & C  & C  & 6 \\ \hline
R. A. Serway and J. W. Jewett
	&C&		NC & C & NC & NT & NT & 2 \\ \hline
R. L. Hawkes, J. Iqbal, \textit{et al.} 
	&C&		NC & C & NC & C & NT  & 3 \\ \hline
H. D. Young and R. A. Freedman
	&C&		NC & C & NC & NT & NT & 2 \\ \hline
R. Wolfson 
 &C&		NC & C & NC & C  & NT & 3 \\ \hline
R. Knight &C&		NC & NC &NC & C  & C  & 3 \\ \hline
\end{tabular} 
\caption{Results corresponding to all criteria.\label{resumen}}
 }
\end{table*}

\section*{Acknowledgment}
The authors would like to thank PEDECIBA (MEC, UdelaR, Uruguay) and
express their gratitude for the grant Fisica Nolineal (ID 722)
Programa Grupos I+D CSIC 2018 (UdelaR, Uruguay). Part of this research
was funded by the Spanish government (MINECO\textbackslash FEDER
PID2019 -105172RB-I00)



\providecommand{\noopsort}[1]{}\providecommand{\singleletter}[1]{#1}%
\end{document}